\documentclass[10pt, conference, compsocconf]{IEEEtran}

\usepackage{graphicx}
\usepackage{multirow}
\usepackage{amsmath}
\usepackage{algorithm}
\usepackage{algorithmic}
\usepackage{subfigure}
\begin{document}

\title{Optimal Number of Cluster Head Selection for\\ Efficient Distribution of Sources in WSNs}

\author{M. S. Fareed, N. Javaid, M. Akbar, S. Rehman$^{\natural}$, U. Qasim$^{\ddag}$, Z. A. Khan$^{\S}$\\

        COMSATS Institute of IT, Islamabad, Pakistan.\\
        $^{\natural}$Iqra University, Islamabad.\\
        $^{\ddag}$University of Alberta, Alberta, Canada.\\
        $^{\S}$Faculty of Engineering, Dalhousie University, Halifax, Canada.
        }

\maketitle

\begin{abstract}
In this paper, we compare problems of cluster formation and cluster-head selection between different protocols for data aggregation and transmission. We focus on two aspects of the problem: (i) how to guess number of clusters required to proficiently consume available sources for a sensor network, and (ii) how to select number of cluster-heads to cover up sensor networks more proficiently. A sensor in Wireless Sensor Networks (WSNs) can communicate directly only with other sensors that are within a radio range in a cluster. However, in order to enable communication between sensors not within communication range, they must form new clusters in distributed sensors. Several clustering algorithms such as LEACH, DEEC, and SEP have been proposed with the objectives of energy minimization, route-path selection, increased connectivity and network longevity. LEACH protocol and the similar ones assume an energy homogeneous system where a node is not likely to fail due to failure in connectivity and packet dropping. More recent protocols like SEP and TEEN considered the reverse that is energy heterogeneity which is more applicable to case of WSNs. We developed a bi-dimensional chain model to select average number of  for DEEC. Simulation results are used to compare performance of different protocols to found optimal solutions of above mentioned problems.
\end{abstract}


\section{Introduction}
Ideally, a cluster-based network can be partitioned into disjoint clusters. Each cluster consists of one Cluster Head (CH) and multiple Member Nodes (MNs). CHs collect data from MNs and relay  processed data to the Base Station (BS). For sake of energy efficiency, it is preferable to create stable and optimal number of clusters, and dynamic CH selection and rotation is desirable over a static CH assignment. Also, CHs are expected to be distributed evenly in network. Therefore, a practical clustering scheme designed for large Wireless Sensor Networks (WSNs) should be distributed and employ dynamic CH selection, cluster formation and periodic CH rotation.

Since the advent of sensor nodes, much work has been done to come up with different models for energy minimization and control. Each of these models has advantages and disadvantages. In general, schemes with more complicated control can lead to (near)-optimal energy efficient solutions. However, this may introduce higher overhead for the coordination and control mechanism, which is also energy and/or time consuming. Therefore, tradeoff should be made. Also, protocols may overlap with each other, i.e. one specific scheme can have the properties in different domains.

Several clustering schemes and algorithms such as Low Energy Adaptive Clustering Hierarchal (LEACH) [1] routing protocol and Distributed Energy Efficient Clustering (DEEC) [3] have been proposed with objectives like fault-tolerance, load balancing, increased connectivity and network longevity. LEACH protocol and similar ones assume an energy homogeneous system where a node is not likely to fail due to failure in connectivity and packet dropping. More recent protocols like Stable Election Protocol (SEP) [2] and Threshold-sensitive Energy Efficient sensor Network (TEEN) [4] considered reverse that is energy heterogeneity which is more applicable to real life scenario for WSNs.

 Background is discussed in next section. Optimal number of CH selection for reducing energy consumption is described in section III. Section IV describes a bi-dimensional chain model to select average number of CH for DEEC. Simulations are perform to compare cluster formation LEACH, SEP, DEEC and TEEN in section V.
\section{Background}
Heinzelman \textit{et al.,} [1] define LEACH, a randomized, distributed clustering protocol, which is widely proposed and tested in WSNs. Much work has been carried out to enhance LEACH. Many flavors of LEACH have cropped up and many authors have tried to develop similar schemes having their basis on this protocol.

In [5], authors tried to obtain an insight into clustering characteristics of distributed, dynamic and randomized DDR [8] clustering schemes, and developed a bi-dimensional Markov chain model to inspect their cluster-forming behavior. Based on proposed analytical model, they also derived formulations of the stochastic properties of LEACH [1], including the distribution of the number of CHs, mean,  standard deviation, and coefficient of variation of number of CHs. This paper was the first effort to make comprehensive classification of clustering schemes in WSNs, to analyze the clustering characteristics of DDR [8] schemes, and address uncertainty problem in cluster formation in WSNs. As a result, a desired number of well-distributed clusters can be created, which leads to higher energy efficiency, better fairness among nodes, and prolonged network lifetime.

An energy efficient cluster ID based routing scheme define in [6]. According to authors, uneven load in network is minimized by cluster size adaptation technique. They gave their routing protocol the name of Cluster ID based Routing in Sensor Networks (CIDRSN) [6], which takes cluster ID as next hop address instead of CH ID in routing table and eliminates cluster formation phase and routing phase from being executed in each round. These phases are only executed during the initialization of network, which reduces the energy consumption and increases the network life to about $16 \%$.

\section{Cluster Formation}
WSNs are an emerging technology. A sensor network system, consists of large number of tiny sensors which use to collect and aggregate data and send it to BS. Sensors are small and low-powered. To establish  WSNs every node sends Hello message to other nodes. In which energy is consumed, to avoid message flooding there are different schemes. Cluster formation is one of them in which whole field is divided into small regions and each region has its CHs. In this way,  work load and energy consumption of nodes is reduced. Once CH is selected, every node in that cluster sends its data to CH, and CH sends it to BS. Cluster architecture decreases opportunity of communication overhearing and power dissipation of sensor nodes. Below few models are discussing pattern of cluster formation.

\subsection{Energy-aware Coverage-preserving Hierarchal Routing (ECHR)}
In [7], a quality of service (QoS) algorithm is specially design for mission critical applications like battle field or health care etc. To guarantee QoS, coverage preservation is an essential task. Yet to maintain sensing coverage there exist a tradeoff with network life time due to limited energy supply of sensor nodes.


ECHR algorithm is design in such a way that it prolongs the network life with a full coverage of points of interests and there is no restriction on BS deployment. BS can be located inside or outside of monitoring area. Transmitting packets through long routes consume more energy and here energy reservation is prime consideration, so first step is to select root node and it is decided by BS, and energy-aware hierarchical routing algorithm is applied to each node. Computation of root node weight of each node ni by:

\begin{eqnarray}
  \alpha_{i}=(q_{i}) ^{\tau_{1}}\times \Bigg( \frac{\|O(s_{i})\|}{\|C(s_{i})\|}\Bigg) ^{\tau_{2}}\times  \Bigg( \frac {1}{d(s_{i},BS)} \Bigg)
\end{eqnarray}

where, $q_{i}$ is residual energy of $s_{i}$, $d(s_{i}, BS)$ is Euclidean distance  between node $s_{i}$ and the BS and $\tau_{1}$ and $\tau_{2}$ are weighting coefficients for residual energy factor and coverage factor respectively.
Root node broadcasts a beacon message in each round. Message contain packet of information that includes its ID, residual energy and level, toward other nodes. First level nodes receive beacon message of root node. Again first level nodes broadcast beacon and nodes that receive this message are second level nodes. In such a way, nodes establish a hierarchical broadcasting and communicate with neighboring nodes.

\subsection{ Cluster ID based Routing in Sensor Networks (CIDRSN)}
In [6], an energy efficient cluster ID based routing scheme is presented. CIDRSN takes cluster ID instead of CHs ID and in this way it balances uneven load in network. In Fig. 1 typical scenario of WSNs, that some nodes sense data and send it to their respective CHS. CH aggregate data and send it to BS through hop by hop communication. In start each node broadcasts a Hello message, which contains node ID and remaining energy of  node. After receiving a predefine number of Hello messages that node broadcasts a message of CH candidate. After receiving this message nodes in  transmission range stop sending Hello messages. As cluster is formed, each CH declares cluster ID which unique. Near BS cluster reduces its size as it has to carry highest network traffic.

\begin{figure*}[!ht]
  \centering
 \subfigure{\includegraphics[height=6 cm, width=18 cm]{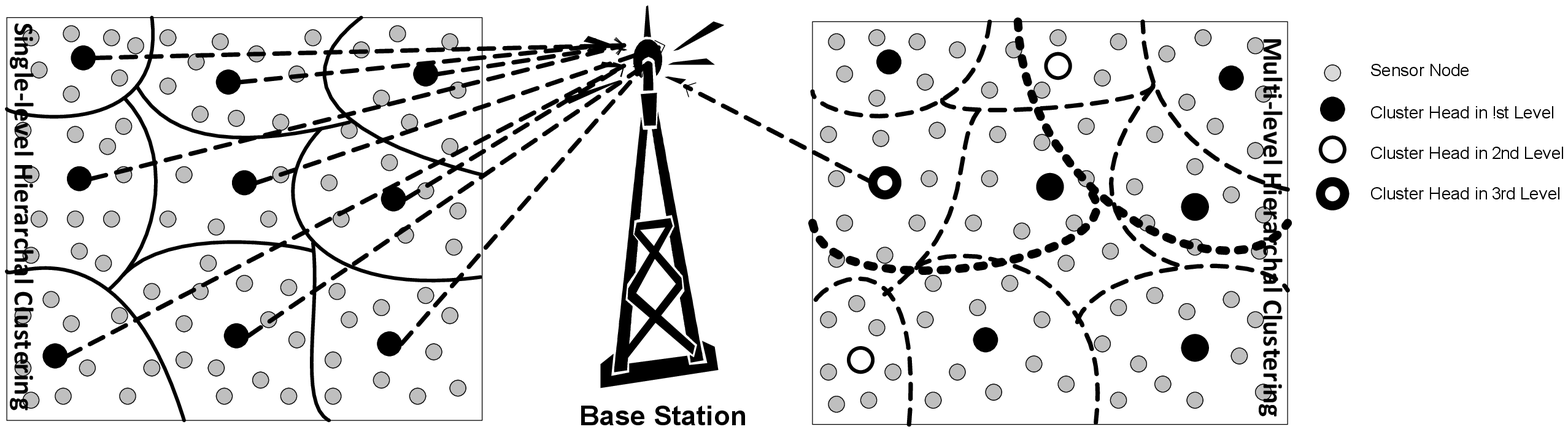}}
  \caption{Multi-level Hierarchal Cluster Based Routing}\label{Fig. 1}
\end{figure*}
\subsection{Estimation of the optimal number of cluster-heads in sensor networks (EONCH)}
LEACH is a clustering scheme in which nodes arrange themselves as a local clusters and each cluster elects CH. Each node in a cluster sends data to CH.  CH process data collected by cluster member nodes and send it to the BS. In this scheme ``Estimation of the optimal number of CHs in sensor network" [5] advertisement as a CH is limited to radio range. CH advertises itself as a CH to sensor nodes within radio range. Sensor nodes within range receive this advertisement and if they are not CH they join  cluster. Now those nodes which are not placed in radio range and don't receive any advertisement from any CH they become forced CHs. Energy utilization in network to gather information from sensor nodes and send it to BS depends on number of CHs and the radio range r of algorithm. During one round each non-CH node sends data to CH once. For calculating optimal number of clusters relation is:

\begin{eqnarray}
  k_{opt}= \Bigg[ \frac{0.5855 N_{\epsilon f s}a^{2}}{\epsilon_{mp}(d*_{toBS})^{4}-E_{elec}} \Bigg]^{1/2}
\end{eqnarray}

\subsection{Probabilistic analysis of Hierarchical cluster protocols for wireless sensor networks}
LEACH is cluster based routing scheme in which energy of  system is managed as transmission range is reduced, each node in cluster send their data to CH. CH forward it to BS. LEACH is round based, every node in cluster becomes CH once in a cycle. In this way energy and load is balanced. Ingemar \textit{et al.,} [8] proposed a probability based model for LEACH in which, selection of CH is based on probabilistic aspects. Each round in LEACH starts with set up phase and steady-state phase. Set up phase selects random number of CH, and form random size of clusters which are of Voronoi type. LEACH is design in such a way that each node become CH once in a cycle. Multi-level LEACH is introduced here which is followed by m-level CH selection algorithm. Selection of probability is based on $\frac{1}{((r_(m-1)-1) )}$. Only $(m - 1)$-CH nodes select a new set of m-CH among  $Xm-1$, $1 - Xm$,$1$ available candidates. Comparison of different attributes of existing protocol with respect to energy, routing and clustering is shown in Table I.

\subsection{LEACH}
Main task in any WSNs is to collect data through sensor nodes and send it to BS. In LEACH [1], we consider that network is homogenous means, all nodes have same energy and they send data to CH which is randomly chosen. Every node become CH once every $1/p$, which is epoch for LEACH. $p$ is probability of becoming CH. Initially each node choose a random number from 0 to 1 and if that number is less than threshold T(s) then that node becomes CH. Calculation of threshold is made by relation given:

\begin{eqnarray}
T (s)=
\begin{cases}\frac{p}{p(r, (mod1/p))}  \;\; if \; s \; \epsilon  \;G \\
0                   \;\;\;\;\;\;\; \;\;\;\;\;\;\; \;\;\;\;\;\;\;otherwise
\end{cases}
\end{eqnarray}

BS is fixed and located far from sensor. CH sends data to BS. LEACH is able to perform local computation in each cluster to reduce the amount of data that must be transmitted to BS.

\subsection{SEP}
SEP [2] is a heterogeneous two-level hierarchical network. As it elects CH with the help of probability and remaining energy in each node. Each node transmits data to closest CH so as to split the communication cost to the sink. In SEP [2] there are two types of nodes normal and advance energy nodes. Advance nodes have more energy and that's why SEP has longer stability period. $E_0$ is the energy of normal node and $E0 (1 + a)$ is energy of advance node. Total initial energy of new two level heterogeneous is:

\begin{eqnarray}
   n . (1 - m) E_0 + n . m. E_0 (1 + a) = n . E_0 . (1 + a . m)
\end{eqnarray}

By introducing heterogeneity energy of system is increased by factor $(1 + a. m)$.  Epoch of the system is $1/ p_{opt} .(1 + a.m)$. In this case system has more energy as compare to LEACH that's why stability period is longer, which is $(1 + a . m)$ times. Threshold relation given in  1 is same for normal and advance nodes.

Sink is located in center of field. Distance of nodes or CH is less than $d0$. In homogenous case every node is equipped with same energy and as first node dies instability period starts. In heterogeneous case advance nodes have more energy than normal. After first node is dead it means that advance node is normal node now. Unstable region is shorter in SEP as compare to LEACH. SEP applies equally well to small sized networks, as well as it is scalable.


\begin{table*}[ht]
 \centering
  \begin{tabular}{| p{3cm} || p{2.5cm} || p{2.5cm} || p{2.5cm} || p{2.5cm} |}
  \multicolumn{5}{c}{Table. I Comparison of attributes of different Routing protocols }\\
  \hline
  \textbf{Routing Protocols}     &\textbf{LEACH}[1] &\textbf{SEP}[2]&\textbf{DEEC}[3]  & \textbf{TEEN}[4]  \\ \hline \hline
  \textbf{Network Type }         & Homogeneous      & Heterogeneous & Heterogeneous    & Heterogeneous \\ \hline
  \textbf{Communication Type}    &   Single-hop     & Single-hop    & Single-hop       & Multi-hop  \\ \hline
  \textbf{Scalability}           &	Limited         &  yes          & yes              &  yes       \\ \hline
  \textbf{Energy-efficiency}	 &  yes             & yes           & yes              &  yes        \\ \hline
  \textbf{Data Aggregation}      &  yes             & yes           & yes              & yes          \\ \hline
  \textbf{CH Selection}&  Residual energy & Residual energy& Residual energy and Average energy  & yes\\ \hline

\end{tabular}
\end{table*}

\subsection{DEEC [3]}
DEEC [3] has heterogeneous-aware clustering algorithm, which prolong network life and stability period. It has multi-level heterogeneous network. In DEEC [3] selection of CH depends on probability which depends on ratio between residual energy of each node and average energy of network. DEEC [3] has rotating epoch depending upon initial and residual energy of node. Total initial energy of system is given as:

\begin{eqnarray}
E_{total}= \sum_{(i=1)}^{N} E_{0}( 1 +a_i )=   E_{0}(N+\sum_{(i=1)}^{N}a_{i})
\end{eqnarray}

Probability of a node to become a CH is $P_{i}$ and nodes with greater energy have larger $P_{i}$ as compare to $P_{opt}$. Let  $\bar{E}$ is average energy of system then we have:

\begin{eqnarray}
 P_i=P_{opt} [1-\frac{(\bar{E}-E_i (r))}{(\bar{E}(r))}]=P_{opt}  \frac{(E_i (r))}{(\bar{E}(r))}
\end{eqnarray}

 Average total number of CH per round per epoch is:

\begin{eqnarray}
\begin{split}
  \sum_{(i=1)}^{N}P_{i}=\sum_{(i=1)}^{N} P_{opt}  \frac{(E_i (r))}{(\bar{E}(r))} \\ =p_opt \sum_{(i=1)}^{N}\frac{(E_i (r))}{(\bar{E}(r))}=N P_{opt }
\end{split}
\end{eqnarray}

Relation for calculating threshold for each round is:

\begin{eqnarray}
T (s)=
\begin{cases}\frac{P_i}{P_i(r, (mod1/P_i))}  \;\; if \; s \; \epsilon  \;G \\
0                   \;\;\;\;\;\;\; \;\;\;\;\;\;\; \;\;\;\;\;\;\;otherwise
\end{cases}
\end{eqnarray}

$E_{poch}$ of system is inverse of $P_i$

\begin{eqnarray}
n_i=\frac{1}{p_i} =\frac{\bar{E} (r)}{(p_{opt} E_i (r)}= n_{opt} \Big( \frac{\bar{E}(r)}{E_i (r)} \Big)
\end{eqnarray}

Note that node which has higher residual energy has more chances to become a CH. Here BS is located in center of a square. All nodes die approximately at same time [3].

In LEACH environment is homogenous and every node has same probability to become CH. When nodes have different energies then network is heterogeneous. We are considering here two cases one is SEP two level heterogeneous and DEEC [3] which is multi level heterogeneous. Weighted probabilities for normal and advance nodes are:

\begin{eqnarray}
P_{nrm}=\frac{p_{opt}}{(1+am)}
\end{eqnarray}

\begin{eqnarray}
P_{adv}=\frac{p_{opt}(1+a)}{(1+am)}
\end{eqnarray}

We extend it to multi-level heterogeneous network then weighted probability $P_{i}$is :

\begin{eqnarray}
P_{i}=
\begin{cases}P_{nrm}=\frac{p_{opt}}{(1+am)} \;\; if \; s_{i} \; is \;Normal \;node \\
P_{adv}=\frac{p_{opt}(1+a)}{(1+am)}         \;\; if \; s_{i} \; is \; Advanced \;node
\end{cases}
\end{eqnarray}

\section{Markov bi-directional}
Clustering is an efficient way in which a network can balance its load, save energy and can enhance network life. Clustering can be of single hop or multi-hop. Different schemes are adopted to prolong network life and for maximum data transmission. For designing any clustering scheme there are two main issues first is how many clusters should be created and second is method of cluster formation. Stochastic modeling of distributed, dynamic, randomized clustering protocol for wireless sensor network scheme is proposed by Wang \textit{et al.,} [5]. In DDR scheme a bi-dimensional Markov chain model is used to study their cluster-forming behavior and derive formulas for statistics of system like probability mass function of number of CHs, average, standard deviation and coefficient of variation of number of CHs. Through DDR scheme uncertainty problems are well handled. Since Markov chain is irreducible and all states are positive recurrent and stationary distribution exists. Through this model we can study uncertainty of number of CHs by observing the distribution and selection of CHs in DEEC [3]. DDR clustering scheme is promising in providing energy efficient, load balancing, scalable and robust communication in WSNs. In DEEC [3] cluster formation and is define as:

\begin{eqnarray}
  P_{(si)}=\frac{P_{(opt)}N(1+a_{i})}{(N+\sum^{N}_{i=1})}
\end{eqnarray}

 where

 \begin{eqnarray}
 P_{i} = P_{opt} \frac{E_{i}(r)}{ \overline{E}(r)}\Longrightarrow P_{(si)}=\frac{P_{(opt)}N(1+a_{i})E_{i}(r)}{(N+\sum^{N}_{i=1})\overline{E}(r)}
 \end{eqnarray}

Number of CHs selected in stage $s_{i}$ has binomial probability

\small
\begin{eqnarray}
f(x)=
 \begin{cases}
P_{(0,N)\longrightarrow(1,0)}= {N \choose i}(P) ^{N-1}(1-P)^{i}    \;\;i\epsilon[0,N] \\
 P_{(0,N)\longrightarrow(1,0)}= {N \choose i}(P) ^{N-1}(1-P)^{i}\\ \;\;\;\;\;\;\;\;P_{(si)}=\frac{P_{(opt)}N(1+a_{i})}{(N+\sum^{N}_{i=1})}\\
 \;\;\;\;\;\;\; \;\;\;\;\;\;\; \;\;\;\;\;\;\;\;\;\; \;\;\;\;\;\;\;i\epsilon[1,N], j\epsilon[0,N]\\
 P_{(m-1,i)}\longrightarrow(0,N)=1 \;\; i\epsilon[0,N]\\
 P_{(S,i)}\longrightarrow(S+1,0)=1 \;\;S\epsilon[1,m-2]
\end{cases}
\end{eqnarray}
\normalsize

First and Second equation in (15)shows number of CHs selected in each stage $s$. Third and fourth equation in (15) deals when all nodes are CHs and no one node is CH respectively. Markov chain rule is defined as:

\begin{eqnarray}
\pi=\pi.P
\end{eqnarray}

where $P$ is transition probability matrix:

\begin{eqnarray}
\begin{cases}
  \pi_{(1,i)}=\pi_{(0,N)}.P_{(0,N)\longrightarrow(1,i)} \;\;i\epsilon[0,N]\\
\pi_{(2,i)}=\sum^{N}_{i=1} \pi_{(1,i)}.P_{(1,i)\longrightarrow(2,i)} \;\;i\epsilon[0,N]\\
... \\                                                                                                                                                                      \pi_{(s,i)}=\sum^{N}_{i_{s-1}=1} \pi_{(s-1,i_{s-1})}.P_{(s-1,i_{s-1})\longrightarrow(s,i)} \\
\;\;\;\;\;\;\; \;\;\;\;\;\;\; \;\;\;\;\;\;\;\;\;\; \;\;\;\;\;\;\; \;\;\;\;\;\;\;\;\;i\epsilon[0,N],S\epsilon[2,m-1]
\end{cases}
 \end{eqnarray}

$F$ is a factor matrix, element $f_{si}, s\epsilon[1,m-1],i\epsilon[0,N]$

\begin{eqnarray}
   \pi_{(s,i)}=\pi_{(0,N)}.P_{(si)} \;\;s\epsilon[1,m-1],i\epsilon[0,N]
\end{eqnarray}
where

\begin{eqnarray}
\begin{cases}
 f_{i} =P_{(0,N)\longrightarrow(i,1)}       \;i\epsilon[0,N]\\
 f_{si}=\sum^{N}_{k=1}f_{(s-1)k}.p_{(s-1)k\longrightarrow(s,i)\;\; s\epsilon[2,m-1],i\epsilon[0,N]}
 \end{cases}
\end{eqnarray}

According to normalization condition:

\begin{eqnarray}
 \pi_{(0,N)}+\sum^{m-1}_{s=1}\sum^{i=1}_{N}\pi_{(s,i)}=1\\
  \pi_{(0,N)}+\sum^{m-1}_{s=1}\sum^{i=1}_{N}\pi_{(0,N)}.f_{si}=1
\end{eqnarray}

\begin{eqnarray}
  \pi_{(0,N)}=\frac{1}{1+\sum^{m-1}_{s=1}\sum^{i=1}_{N}.f_{si}}
\end{eqnarray}

Using induction property we have :

\begin{eqnarray}
 \sum^{N}_{i=0} f_{si}=1, \; s\epsilon[1,m-1]
\end{eqnarray}

Putting Eq. (22) in (23) we have:

\begin{eqnarray}
 \pi_{(0,N)}=\frac{1}{1+m-1}=P
\end{eqnarray}

Suppose $ch$ denote a random variable representing number CHs in a round. Then by transition probabilities and Markov chain model the $pmf$ of the number of CHs is:

\begin{eqnarray}
\begin{split}
  p(ch=k)=\pi_{(0,N)}.P _{(0,N)\rightarrow (1, N-k)}+\pi_{(m-1,k)}+ \\ \sum^{m-2}_{s=1}\sum^{N}_{i=k}\pi_{(s,i)}P_{(s,i)\rightarrow(s+1,i-k)}
\end{split}
\end{eqnarray}

Average number of CHs is :
\begin{eqnarray}
  ave[ch]=\sum^{N}_{k=0}.P(ch=k)
\end{eqnarray}

\section{Simulation Results}
We perform simulations to compare performance of selected protocol for cluster formation and CHs selection. We use MATLAB as a simulator to analyze performance of cluster base routing protocol. We take network size of $100m$ x $100m$ in which $100$ nodes are randomly distributed. BS is placed in any arbitrary position. All parameters taken for these simulations are defined in Table. II.

\vspace{0.5cm}
\begin{table}[!ht]
\begin{center}
  \begin{tabular}{| p{3cm} || p{3cm} |}
  \multicolumn{2}{c}{Table. II Simulation Environment}\\
  \hline
  \textbf{Parameters} & \textbf{Value}   \\ \hline \hline
   \textbf{Size of Network	}&  100 m x 100 m	 \\ \hline
    \textbf{$E_{elec}$ (Radio Electronics Energy)}	&  50 nJ/bit	 \\ \hline
     \textbf{$E_{amp}$ (Radio Amplifier Energy )}& 100 pJ/bit/$m^{2}$	 \\ \hline
      $E_{fs}$ & 10 pJ/bit/$m^{2}$	 \\ \hline
      $E_{initial}$ \textbf{(Intial Energy)}	&  0.5 J	 \\ \hline
      \textbf{Number of Nodes}& 100	 \\ \hline
      $E_{DA}$	& 5 nJ/bit/message	 \\ \hline
      $P_{opt}$	& 0.1, 0.2, 0.3	 \\ \hline
\end{tabular}
\end{center}
\end{table}

CHs in each round are selected by distributed algorithm. Optimum number of CH selected is restrictively depends on number of sensors nodes in entire sensor network.  Fig. 2 (a, b)  and (c) show how different number of clusters and cluster sizes varies with number of rounds. Also number of clusters is relative to sensor network size. In other words, the larger the sensor network, more clusters are required to apply optimal data aggregation. In beginning of network more CHs are required to aggregate data. However, with passage of time less CHs are required. Since network size is decreasing due to sensors are dying.  DEEC [3] selects more CHs as compared to TEEN [4], SEP [2] and LEACH [1]. DEEC [3] CHs selection almost remain around $10-60$ in each round.
\begin{figure}[!ht]
  \begin{flushright}
 \subfigure[When 10\% nodes are Cluster Heads]{\includegraphics[height=5 cm, width=8 cm]{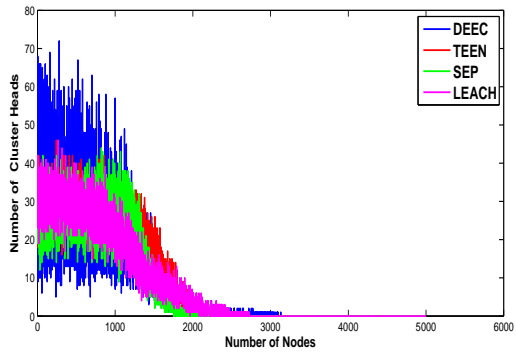}}
\subfigure[When 20\% nodes are Cluster Heads]{\includegraphics[height=5 cm, width=8 cm]{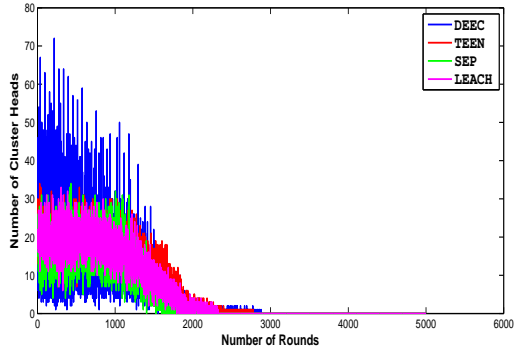}}
\subfigure[When 30\% nodes are Cluster Heads]{\includegraphics[height=5 cm, width=8 cm]{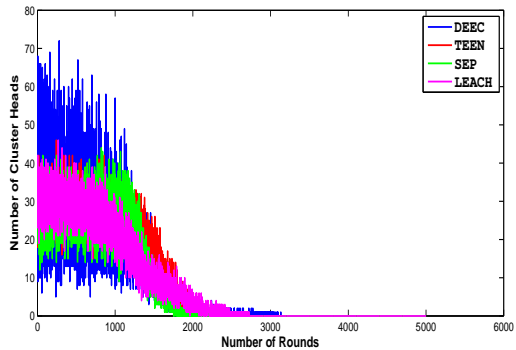}}
  \caption{Optimal number of Cluster Heads Selection}\label{Fig. 4}
   \end{flushright}
\end{figure}

\section{Conclusion}
WSNs are networks of light-weight sensors that are battery powered used primarily for monitoring purposes. The advances in micro-electromechanical technologies have made the development of such sensors a possibility. While WSNs are increasingly equipped to handle complex functions such as data aggregation, information fusion, computation and transmission activities, these sensors require to use their energy efficiently to extend their effective network life time. Since sensor nodes are prone to energy drainage and failure, thus constant re-energizing is required as old sensor nodes die out. This can damage  the stability and performance of the network system if energy is not properly utilized. MATLAB simulations are performed to compare cluster formation of LEACH [1], SEP [2], DEEC [3] and TEEN [4]. From the Simulation results it is found that DEEC [3] selects optimal number of CHs to forward data to BS and to increase life time of network.

\end{document}